\documentclass[eqsecnum,aps,showpacs,twocolumn,floatfix]{revtex4} 
\usepackage{graphicx}

\begin{document}
   
 
\title{Dynamics of collapsing and exploding 
Bose-Einstein condensed vortex  state}
 
\author{Sadhan K. Adhikari}
\affiliation{Instituto de F\'{\i}sica Te\'orica, Universidade Estadual
Paulista, 01.405-900 S\~ao Paulo, S\~ao Paulo, Brazil\\}

\date{\today}

\begin{abstract}

Using the time-dependent mean-field Gross-Pitaevskii equation we study the
dynamics of small repulsive Bose-Einstein condensed vortex states of
$^{85}$Rb atoms in a cylindrical trap with low angular momentum $\hbar L$
per atom $(L\le 6)$, when the atomic interaction is suddenly turned
attractive by manipulating the external magnetic field near a Feshbach
resonance.  Consequently, the condensate collapses and ejects atoms via
explosion and a remnant condensate with a smaller number of atoms emerges
that survives for a long time. Detail of this collapse and explosion is
compared critically with a similar experiment performed with zero angular
momentum ($L=0$). Suggestion for future experiment with vortex state is
made.

\end{abstract}
\pacs{03.75.Fi}

\maketitle
 

\section{Introduction}
 
Recent observation \cite{1,ex2} of Bose-Einstein condensates (BEC) of
dilute trapped bosonic atoms with repulsive interaction at ultra-low
temperature has intensified theoretical and experimental studies on
various aspects of the condensate \cite{11}.
 One fascinating feature is the observation of quantized vortices
\cite{exp} and
vortex lattice \cite{exp1,expm}  in the
condensate as this is intrinsically related to the existence of
superfluidity.  Another interesting feature is the formation of a stable condensate
composed of a finite number of attractive atoms less than a critical
number $N_{\mbox {cr}}$ \cite{ex2}.  The third  noteworthy feature is
the
observation of Feshbach resonances in $^{23}$Na \cite{fbna}, $^{85}$Rb
\cite{ex3} and $ $Cs \cite{fbcs} atoms, as in the presence of such
resonances the effective atomic interaction can be varied in a controlled
fashion by  an external (background) magnetic field \cite{fbth}.

For superfluid $^4$He(II) in a rotating container, no motion of the fluid
is observed below a critical rotational frequency. Above this
frequency quantized vortices appear in  $^4$He(II) manifesting its
superfluidity.   However, because of
the strong interaction, the
theoretical description of this system is not easy. Quantized
vortices have been observed \cite{exp,exp1,expm}
in trapped BEC  
and 
can be generated in theoretical mean-field models
\cite{2a,2c,2d,2f,2ff,2g,fc}
based on the Gross-Pitaevskii (GP)  \cite{8} equation.
Different ways
for generating  vortices in a BEC have been suggested 
\cite{2d},
e.g., via spontaneous formation in evaporative cooling \cite{2f},
via controlled excitation to an excited state \cite{2ff}, by
stirring a BEC using a  laser with an angular frequency above a
critical value  
\cite{2c}, or by the rotation of an axially symmetric trap with an angular
frequency above a similar critical value \cite{2g}.
In contrast to liquid $^4$He(II), a trapped BEC of small size
is
dilute and weakly interacting, which makes a mean-field
analysis appropriate.

The observation of a condensate of attractive $^7$Li atoms and the
subsequent measurement of the critical number $N_{\mbox{cr}}$ \cite{ex2}
is in good agreement with the mean field analyses in a spherically
symmetric trap \cite{skxx}, although the agreement is not as good in the
case of $^{85}$Rb \cite{unnum} atoms in an axially symmetric trap
\cite{sk1}. If the number of atoms can somehow be increased beyond
this critical number, due to interatomic attraction the condensate
collapses emitting atoms until the number of atoms is reduced below
$N_{\mbox{cr}}$. With a 
supply of atoms from an external source the condensate grows again beyond
$N_{\mbox{cr}}$ and a sequence of collapses has been observed in
$^7$Li by Gerton et al at Rice \cite{ex2}, where the number of atoms
remain close to $N_{\mbox{cr}}$ and the collapse is driven by a stochastic
process.

Recently, 
a more challenging experiment was performed by Donley et al.
\cite{ex4} at JILA on a BEC of $^{85}$Rb atoms
\cite{ex3} in an axially symmetric trap, where they varied the
interatomic interaction by an  external magnetic field near a
Feshbach resonance \cite{fbth}.  Consequently, they were able to change
the
sign of the atomic scattering length, thus transforming a repulsive
condensate of $^{85}$Rb atoms into a collapsing and highly explosive
attractive condensate and studied the dynamics of the same \cite{ex4}.  
Immediately after the jump in the scattering length, one has a highly
unstable BEC, where the number of atoms could be much larger than
$N_{\mbox{cr}}$.  Donley et al. \cite{ex4} have provided a quantitative
estimate of
the explosion  by measuring the number of atoms remaining in
the BEC as a function of time until an equilibrium is reached.  
Because this phenomenon of emission of a very large number of atoms in a
small interval of time is reminiscent of an explosion and 
looks very much like a tiny supernova, or exploding
star, the researchers dubbed it a ``Bosenova".

The essential aspects of the above experiments at Rice by Gerton et al
\cite{ex2} and
at JILA by Donley et al \cite{ex4} have been theoretically described by a
variety of
authors using the GP equation \cite{8}.  The
theoretical analyses have not only produced the time-independent results,
such as, the critical number $N_{\mbox{cr}}$ \cite{skxx,sk1}, but
also
time-dependent
results, such as, the variation of number of atoms of the BEC during
collapse and explosion \cite{th1,th1a,th2,th3,th4,th41}, both in
reasonable
agreement with
experiment. This
consolidates the use of the mean-field GP equation in describing the
dynamics of
collapsing and exploding BEC of small to medium size. These BEC's composed
of several thousand atoms can be considered  dilute and weakly
interacting and hence amenable to mean-field treatment.  Motivated by the
above success, using the GP equation  we propose the numerical simulation 
of the dynamics of a rotating Bosenova with a single vortex
composed of a small number (several thousands) of $^{85}$Rb
atoms as in the experiment at JILA \cite{ex4}.   
We consider a
single vortex state \cite{exp,2a,sk1}, as appropriate for small
condensates, opposed to
vortex lattice for large condensates \cite{exp1,expm,fc}. 
A comparison of the
present results with future experiment will provide a more stringent test
for  the mean-field GP equation.

In this paper
we perform a mean-field analysis based on the time-dependent
GP equation to understand  the  collapse and
explosion of the attractive vortex state  of $^{85}$Rb atoms in an
axially symmetric trap. 
To account for the loss of atoms from the
strongly attractive condensate we include an absorptive
nonlinear three-body recombination term in the GP equation. The
three-body recombination rate we use in numerical  simulation
is the same as used in a similar study with BEC's with zero angular
momentum \cite{th4} and is 
in agreement with previous
experimental measurement \cite{k3} and theoretical calculation
\cite{esry}.
In the present investigation we consider the complete numerical solution
of the mean-field GP equation for an axially symmetric trap as in the
experiment at JILA \cite{ex4}. As in the experiment at JILA with
nonrotating condensates,
we find that,  also 
in the case of rotating vortex states,  a large number of atoms could be
emitted in a small interval of time and one could have a Bosenova-type
explosion.

Throughout the present numerical simulation we make the assumption
that the axial symmetry of the system is maintained. For small values of
nonlinearity  a dynamical quadrupole instability may cause an
attractive BEC vortex state to split into two pieces that rotate around
the axial direction \cite{ref1}.  These pieces may unite to recover the original vortex
and this split-merge cycle repeats.  Similar instability is known to exist
for BEC in a toroidal trap \cite{ref2}.  Clearly, a full three-dimensional
calculation of the collapsing phenomena taking into consideration the
effect of the splitting of vortex states seems to be practically
impossible at present and will be a welcome future work.  In view of this
here we present an axisymmetric model of the same which is expected to
provide the essentials of the collapse dynamics of the vortex states.

In Sec. II we present the theoretical model and the numerical method for
its solution. In Sec, III we present our results. 
Finally, in Sec. IV we present a brief
discussion and concluding remarks.
 
\section{The Gross-Pitaevskii Equation}
 
\subsection{Theoretical Model}
 
The time-dependent Bose-Einstein condensate wave
function $\Psi({\bf r};\tau)$ at position ${\bf r}$ and time $\tau $
allowing
for atomic loss
may
be described by the following  mean-field nonlinear GP equation
\cite{11,8}
\begin{eqnarray}\label{a} \biggr[& -& i\hbar\frac{\partial
}{\partial \tau}
-\frac{\hbar^2\nabla^2   }{2m}
+ V({\bf r})
+ {\cal G}N|\Psi({\bf
r};\tau)|^2-  \frac{i\hbar}{2}
 (K_2N \nonumber \\
& \times &|\Psi({\bf r};\tau) |^2
+K_3N^2|\Psi({\bf r};\tau) |^4)
 \biggr]\Psi({\bf r};\tau)=0.
\end{eqnarray}
Here $m$
is
the mass and  $N$ the number of atoms in the
condensate,
 ${\cal G}=4\pi \hbar^2 a/m $ the strength of interatomic interaction, with
$a$ the atomic scattering length.  
The terms $K_2$ and $K_3$ denote two-body
dipolar and three-body recombination loss-rate coefficients, respectively.
There are many ways to account for the loss mechanism \cite{th1,th1a}. 
Here we
simulate the atom  loss via
the most important quintic three-body term  $K_3$ \cite{th1,th2,th3}.
The contribution of the cubic  two-body  loss term
\cite{k3} is
expected to be negligible \cite{th1,th3} compared to the  three-body term
in
the present problem of the  collapsed condensate with large density
and will not be considered here.

The trap potential with cylindrical symmetry may be written as  $  V({\bf
r}) =\frac{1}{2}m \omega ^2(r^2+\lambda^2 z^2)$ where
 $\omega$ is the angular frequency
in the radial direction $r$ and
$\lambda \omega$ that in  the
axial direction $z$, with $\lambda$ the aspect ratio. We are using the
cylindrical
coordinate system ${\bf r}\equiv (r,\theta,z)$ with $\theta$ the azimuthal
angle.
The normalization condition of the wave
function is
$ \int d{\bf r} |\Psi({\bf r};\tau)|^2 = 1. $

The GP equation can easily accommodate quantized vortex states with
rotational motion of the BEC around the $z$ axis. In such a vortex the
atoms flow with tangential velocity $L\hbar/(mr)$ such that each atom has  
quantized angular momentum $L\hbar$ along the $z$ axis. This corresponds
to an angular dependence of 
\begin{equation}\label{aa}
\Psi({\bf r}, \tau)= \psi(r,z,\tau)\exp (iL\theta)
 \end{equation}
of the wave function, where $\exp (iL\theta)      $ are the circular
harmonics in two dimensions. 
 
Now  transforming to
dimensionless variables
defined by $x =\sqrt 2 r/l$,  $y=\sqrt 2 z/l$,   $t=\tau \omega, $
$l\equiv \sqrt {\hbar/(m\omega)}$,
and
\begin{equation}\label{wf}
\frac{ \varphi(x,y;t)}{x} \equiv   \sqrt{\frac{l^3}{\sqrt 8}}\psi(r,z;\tau),
\end{equation}
we get from Eqs. (\ref{a}) and  (\ref{aa})
\begin{eqnarray}\label{d1}
&\biggr[&-i\frac{\partial
}{\partial t} -\frac{\partial^2}{\partial
x^2}+\frac{1}{x}\frac{\partial}{\partial x} -\frac{\partial^2}{\partial
y^2}+{L^2-1\over x^2}
+\frac{1}{4}\left(x^2+\lambda^2 y^2\right) \nonumber \\
&+& \kappa   n\left|\frac {\varphi({x,y};t)}{x}\right|^2
- i\xi n^2\left|\frac {\varphi({x,y};t)}{x}\right|^4
 \biggr]\varphi({ x,y};t)=0, 
\end{eqnarray}
where
$ n =   N a /l$, $\kappa = 8\sqrt 2 \pi$, 
and $\xi=4K_3/(a^2l^4\omega).$
From theoretical \cite{ver} and experimental \cite{k3} studies
it has been found that for negative $a,$ $K_3$ increases rapidly as
$|a|^n$, where the theoretical study 
favors $n=2$ and we represent this variation via this  quadratic
dependence. This makes the parameter $\xi$ above a constant \cite{th4} for
an experimental set up with fixed $l$ and $\omega$
and in the present study we  employ a
constant $\xi$.

The normalization condition  of the wave
function  for  $K_3=0$ is 
\begin{equation}\label{5} {\cal N}_{\mbox{norm}}\equiv {2\pi} \int_0
^\infty
\frac{dx}{x} \int _{-\infty}^\infty dy|\varphi(x,y;t)|
^2 =1.  \end{equation}
However, in the presence of loss
$K_3 > 0$, ${\cal N}_{\mbox{norm}}
< 1.$ The number of remaining atoms $N$
in the condensate is given by $ N=N_0
{\cal N}_{\mbox{norm}}$,
 where $N_0$ is the initial number.
 
The root mean square (rms) sizes  $x_{\mbox{rms}}$ and  $y_{\mbox{rms}}$
are
defined by
\begin{eqnarray}
x^2_{\mbox{rms}}= \frac{2\pi}{{\cal N}_{\mbox{norm}}} \int_0
^\infty
dx \int _{-\infty}^\infty dy|\varphi(x,y;t)|
^2 x,   \\
y^2_{\mbox{rms}}=  \frac{2\pi}{{\cal N}_{\mbox{norm}}} 
 \int_0
^\infty
\frac{dx}{x} \int _{-\infty}^\infty dy|\varphi(x,y;t)|
^2 y^2.
\end{eqnarray}

\subsection{Calculational  Detail}

We solve the GP equation (\ref{d1}) numerically  using a
split-step time-iteration
method
using the Crank-Nicholson discretization scheme 
\cite{sk1,th4,sk2,sk3,koo}.  We discretize the GP equation with time step 0.001
and space step 0.1 spanning $x$ from 0 to 15 and $y$ from $-35$  to 35. 
 
It is now appropriate to calculate the parameters of the present
dimensionless GP equation (\ref{d1}) corresponding to the experiment at
JILA for $L=0$ \cite{ex4}. As in that experiment we take 
the radial and axial trap frequencies to be  $\nu_{\mbox{radial}}=17.5$ Hz
and  $ \nu_{\mbox{axial}}=6.8$ Hz,
respectively, leading to $\lambda = 0.389 $. The harmonic oscillator
length $l$ of  $^{85}$Rb atoms for $\omega =2\pi\times 17.5$ Hz and 
$m\approx 79176$
MeV
is
$l=\sqrt{\hbar/(m\omega)}=26070$ \AA. One unit of time $t$ of
Eq. (\ref{d1}) is $1/\omega$ or 0.009095 s.

We consider  a stable $^{85}$Rb
condensate of $N_0= 16000$ atoms with scattering  length
$a_{\mbox{initial}}=7a_0$,
$a_0=0.5292$ \AA. This wave function is obtained 
by time iteration of Eq. (\ref{d1}) employing  the following normalized 
initial solution with a single central vortex
\cite{sk1}
\begin{equation}
\varphi(x,y)=\left[{\lambda \over 2^{2L+3}\pi^3(|L|!)^2}  \right]^{1/4}
x^{1+|L|}e^{-(x^2+\lambda y ^2)/4} 
\end{equation}
for $n=0$. 
In the course of  above time iteration the nonlinear parameter $n$ was 
increased by steps of 0.0001 until its  final value is attained. 
Then during an
interval of time 0.1 ms the scattering length was ramped to  $a=
a_{\mbox{collapse}}$.  The absorptive term $\xi$ was set equal
to zero
throughout  above time iteration.

The final condensate is strongly attractive
and unstable and undergoes a sequence of collapse and explosion. In our
numerical simulation with $L\ne 0$ we consider 
a set of different values of  $a_{\mbox{collapse}}$ $ (=
-263a_0, -100a_0, -30a_0, -20a_0,$ etc.)
 as well as $N_0=6000$.

For the simulation of collapse and explosion a nonzero value of $\xi$ $
(=2)$ is chosen for different $a_{\mbox{initial}}$, $a_{\mbox{collapse}}$,
and $N_0$ as in Ref.  \cite{th4} and the time-evolution of the GP equation
is continued. This value of $\xi$ reproduced the essentials of the
experiment at JILA \cite{ex4} reasonably well for $L=0$ as well as
produced \cite{th4} a $K_3$ in reasonable agreement with a previous
experiment ($K_3=4.24\times 10^{-25}$ cm$^6$/s) \cite{k3} and theoretical
calculation
($K_3=6.7\times 10^{-25}$ cm$^6$/s) \cite{esry} for $a = -370
a_0$. In particular we use $K_3=9\times 10^{-25}$ cm$^6$/s for $a = -370
a_0$. For smaller values of $|a|$, the $K_3$ values are scaled down using
the relation $K_3 \propto a^2$.

\section{Numerical Result}

\begin{figure}[!ht]
 
\begin{center}
\includegraphics[width=\linewidth]{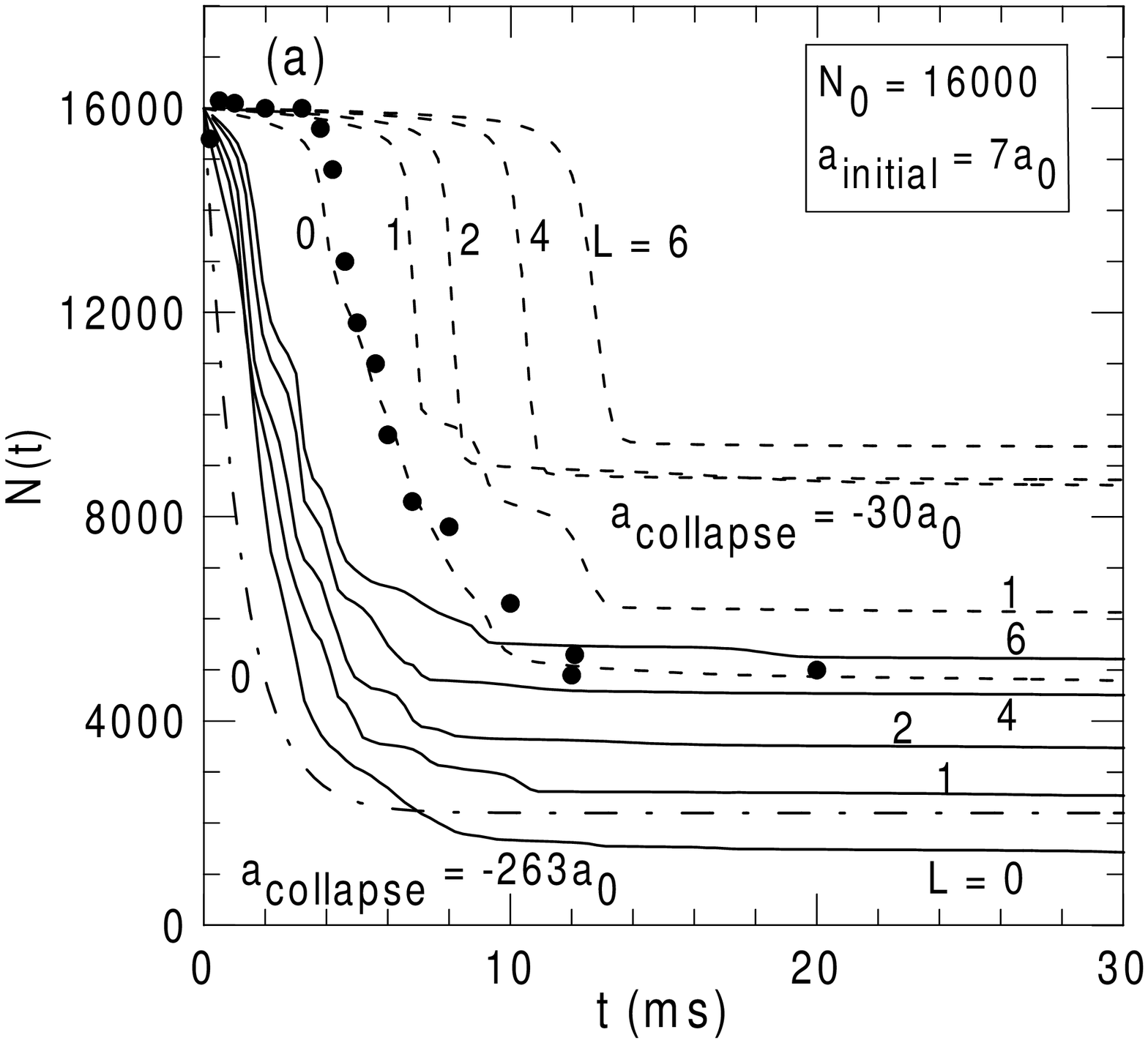}
\includegraphics[width=\linewidth]{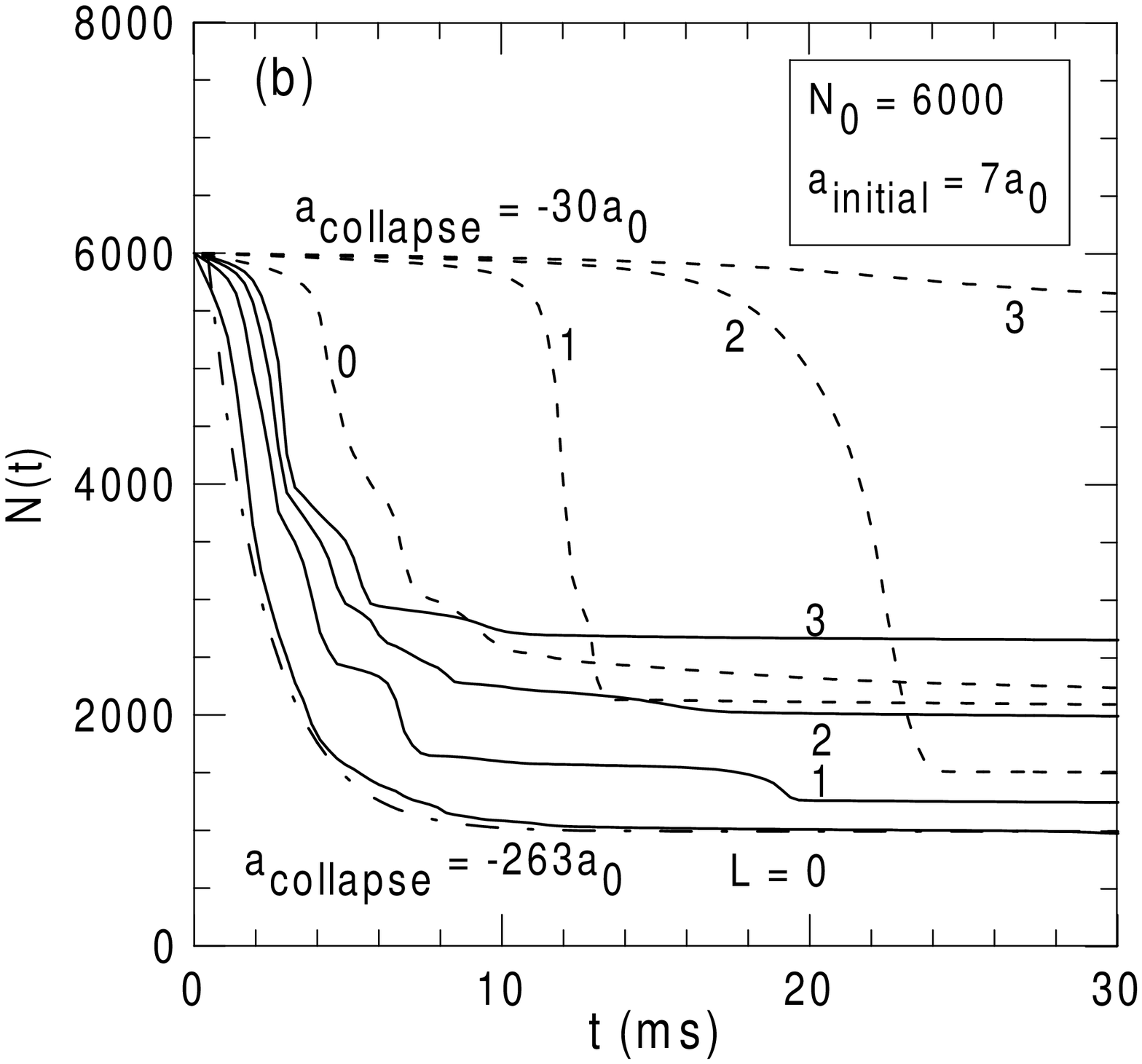}
\end{center}
 
\caption{The number of remaining atoms $N(t)$ in the condensate of
(a) $N_0=16000$ and (b) 6000 atoms after ramping the scattering length
from $a_{\mbox{initial}}=7a_0$ to $a_{\mbox{collapse}} = -30a_0$
(dashed line) and
$-263a_0$  (full line) for different $L$ as a function of time $t$. The
curves are
labeled by their respective $L$ values. Solid circles represent
results of experiment at JILA \cite{ex4} for $L=0$, $N_0=16000$  and
$a_{\mbox{collapse}} = -30a_0$ and dashed-dotted lines represent the
average  \cite{th4} over experimental results at  JILA \cite{ex4} for
$L=0$ and
$a_{\mbox{collapse}} = -263a_0$}
 
\end{figure}

The numerical simulation using Eq. (\ref{d1}) with a nonzero $\xi $ as
described above  immediately yields the remaining number of atoms in the
condensate after the jump in scattering length.
The remaining  number of atoms vs. time is plotted in Fig. 1 (a)  for
$a_{\mbox{initial}}=7a_0$, $a_{\mbox{collapse}}=-30a_0$ and $-263a_0$,
$N_0=16000$, and $L=0,1,2,4,$ and 6. 
In Fig. 1 (b) the same results for 
$N_0=6000$ are plotted.
In this figure we also plot some 
results of experiment at JILA for $L=0$ \cite{ex4,th4}. These experimental
results are in
agreement
with the
simulation for $L=0$. In Fig. 2 (a)  we plot
the particle loss curves for $N_0=6000$,  
$a_{\mbox{initial}}=7a_0$ and different  $a_{\mbox{collapse}}$ for 
$L=0$. In Fig. 2 (b) we plot the same for $L=1$.

From Figs. 1 and 2 we find that 
for a fixed $N_0$,  for a
sufficiently small $L$ or a
sufficiently large $|a_{\mbox{collapse}}|$, there could be
collapse and explosion  during a relatively short interval of time (called decay
time) with the loss of a large fraction of the atoms. However, there is no
collapse for a large enough $L$ or a small  enough $|a_{\mbox{collapse}}|$. For
example, for $N_0 =6000$ in Fig. 1 (b) there is no collapse for $L>2$ for
$|a_{\mbox{collapse}}|=30a_0$
and in
Fig. 2 (a)  there is
no collapse for  $|a_{\mbox{collapse}}|< 5 a_0$ for $L=0$.

\begin{figure}[!ht]
 
\begin{center}
\includegraphics[width=\linewidth]{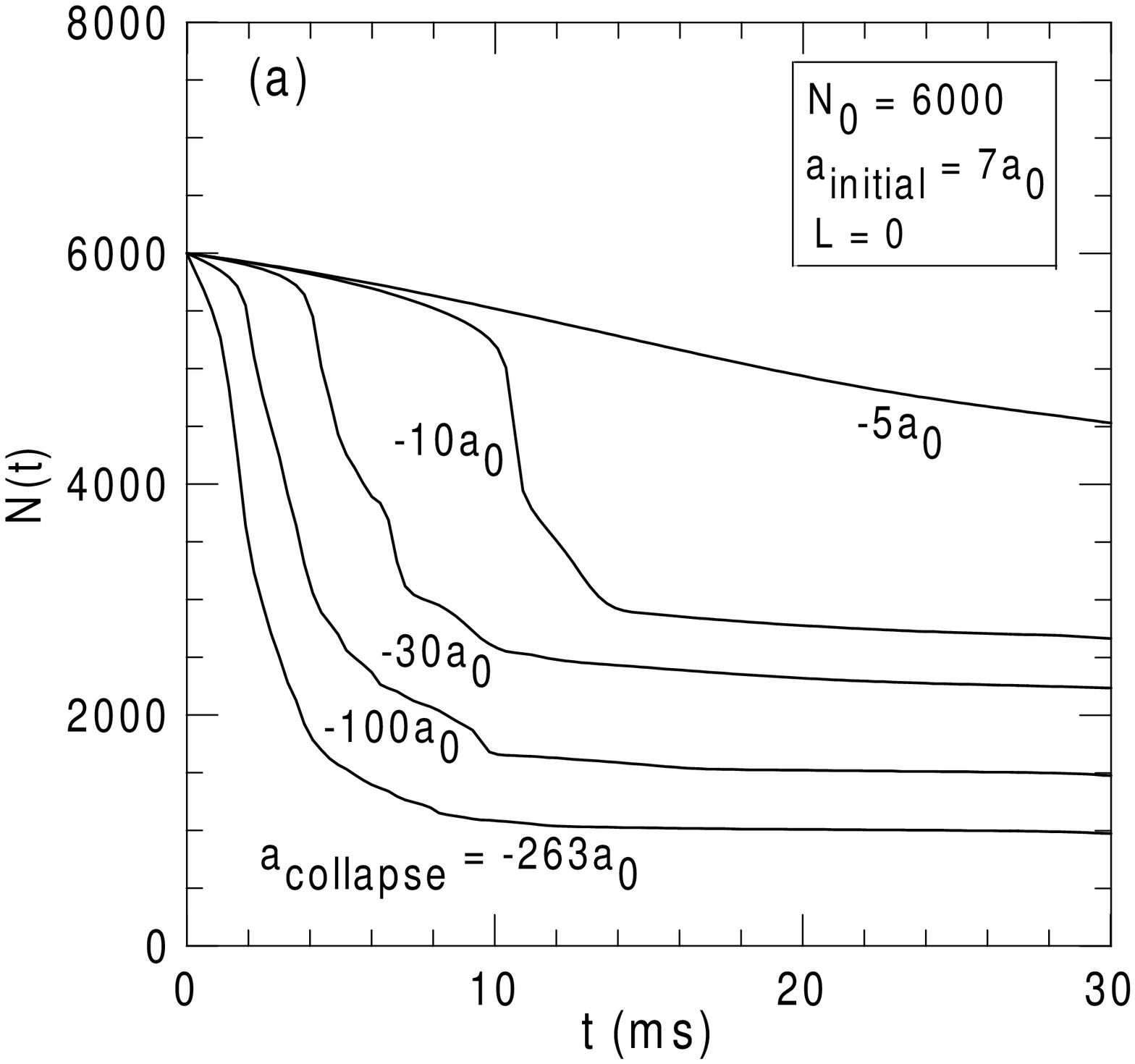}
\includegraphics[width=\linewidth]{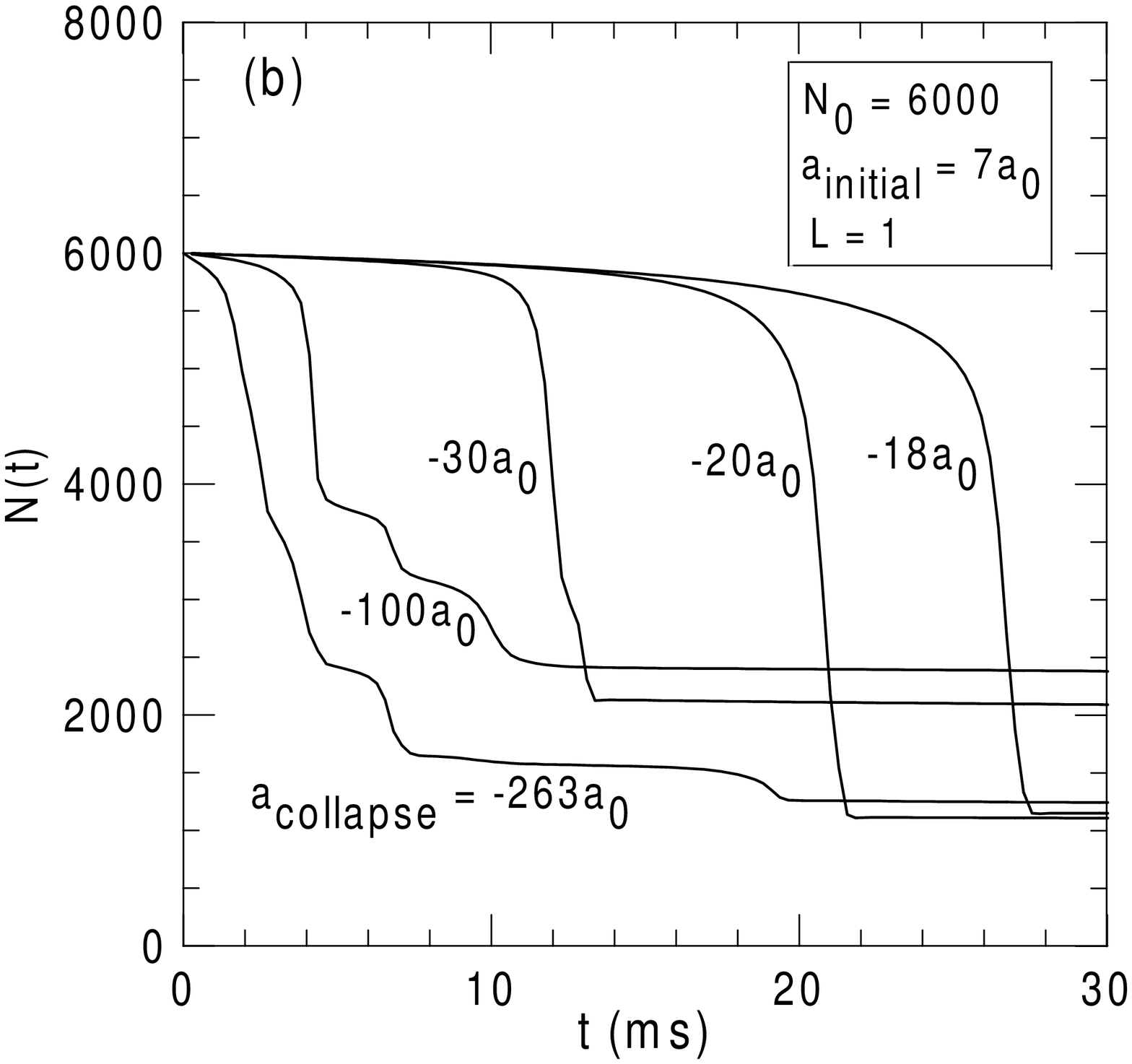}
\end{center}

\caption{The number of remaining atoms $N(t)$ in the condensate of
$N_0=6000$  atoms for (a) $L=0$ and (b) $L=1$  after ramping the
scattering length
from $a_{\mbox{initial}}=7a_0$ to different final $a_{\mbox{collapse}}$
as
a function of time $t$. The
curves are
labeled by their respective  $a_{\mbox{collapse}}$    values.}

\end{figure}

\begin{figure}[!ht]
 
\begin{center}
\includegraphics[width=\linewidth]{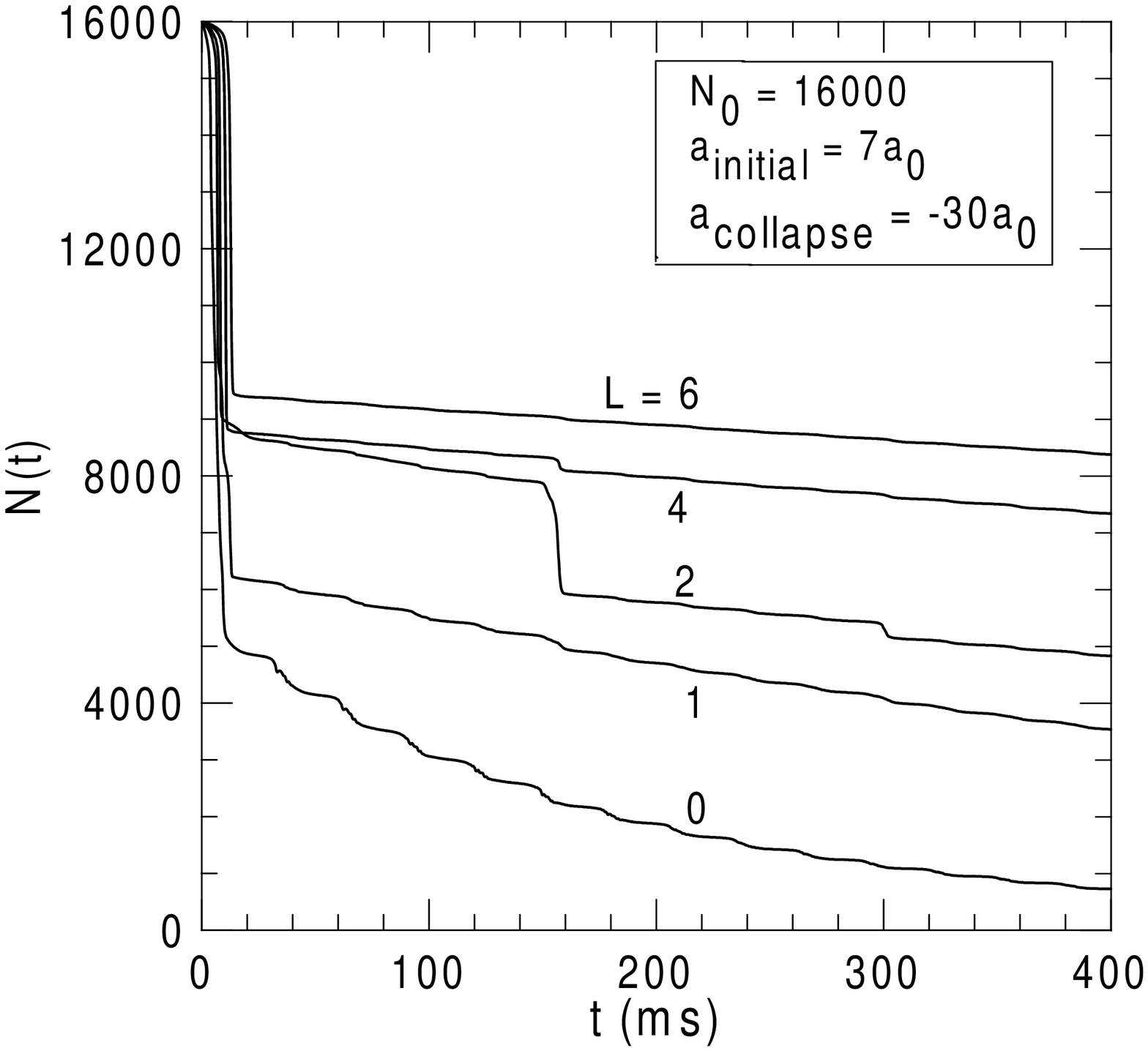}
\end{center}

\caption{The number of remaining atoms $N(t)$ in the condensate of
$N_0=16000$ atoms for different $L$   after
ramping the
scattering length
from $a_{\mbox{initial}}=7a_0$ to the final $a_{\mbox{collapse}}=-30a_0$
as
a function of time $t$. The
curves are
labeled by their respective  $L $    values.}
\end{figure}

In the experiment at JILA \cite{ex4} for $L=0$ it was observed that the
strongly attractive condensate after preparation remains stable with a
constant number of atoms for an interval of time $t_{\mbox{collapse}}$,
called collapse time. This behavior is physically expected for medium to
small values of $|a_{\mbox{collapse}}|$  $ (< 50a_0)$. Immediately after
the
jump in the scattering length to a negative value, the attractive
condensate
shrinks in size during $t_{\mbox{collapse}}$, until the central density
increases to a maximum. Then the absorptive three-body term takes full
control to initiate the explosion that last for few
milliseconds. Consequently, the number of atoms
remains constant for time $t<t_{\mbox{collapse}}$.  The
present results in Figs. 1(a) and (b)  also show this behavior for
$|a_{\mbox{collapse}}|=30a_0$. However, for larger $|a_{\mbox{collapse}}|$
$(=263a_0)$, the atomic attraction is very strong and the central density
increases to a maximum quickly to start the explosion and
$t_{\mbox{collapse}}$ is close to zero. 
In Figs. 2 (a) and (b)  we 
see the dependence of particle
loss and $t_{\mbox{collapse}}$ on $|a_{\mbox{collapse}}|$  
for $N_0=6000$,
$a_{\mbox{initial}}=7a_0$, and  
$L=0$ and $L=1$, respectively. From Figs. 1 and 2 we find that
$t_{\mbox{collapse}}$ increases with $L$ for a fixed
$a_{\mbox{collapse}}$ and with 
$a_{\mbox{collapse}}$ for a fixed $L$. 

From Figs. 1 and 2 we find that after the collapse, the number of
particles drop sharply during a small interval of time called decay time
(few milliseconds),
which means that
the condensate emits a large number particles in an explosive fashion. 
This emission of particles is termed explosion.

After a sequence of collapse and explosion, for $L=0$ 
Donley et al. \cite{ex4}  observed a ``remnant" condensate of
$N_{\mbox{remnant}}$ atoms at large times containing a 
fraction of the initial $N_0$ atoms. Figures 1 and 2  show such a
behavior for different values of $L$ and 
$a_{\mbox{collapse}}$.  In all cases the decay time during which the explosion
takes place is small and of the order of few milliseconds. The decay time
for vortex states ($L\ne 0$)  is smaller than for nonrotating condensates
$(L=0)$. 

We studied the time evolution of the condensate for larger times.  In Fig.
3 we plot the loss curves for $N_0 =16000$, $a_{\mbox{initial}}=7a_0$,
$a_{\mbox{collapse}}=-30a_0$, and $L=0,1,2,4,6$ at larger times.  The BEC
continues to lose atoms if left for a long time but at a rate much slower
than during the first explosion that we call primary.  However, in the
experiment at JILA Donley et al observed that a remnant condensate
containing a fraction of the atoms survived with nearly constant number
for more than one second.  One possible reason for this discrepancy could
be the following.  In the actual experiment at JILA a major portion of the
emitted atoms, called the burst atoms, remain trapped and oscillate around
the central remnant. The presence of the  
burst atoms
make the measurement of the number of atoms in the remnant a difficult
task \cite{don}.
Some of these burst atoms may also rejoin the remnant
to compensate for the three-body loss at large times. 
Such an effect is
not included in the present model, which, hence, presents a larger loss
for the remnant compared to experiment.

We
also observe an interesting phenomenon in Fig. 3, e. g., the
occurrence of smaller secondary and tertiary explosions after the primary
one observed for small times.  For $L=2$ after the primary collapse and
explosion at $t<10$ ms with the loss of about 7000 atoms, there is another
collapse and explosion with loss of about 2000 atoms at $t \approx 150$ ms.  
The primary and secondary explosions are separated by a large interval of
time. We also see much weaker explosion(s) in the course of time evolution
in Fig. 3, where the particle number varies in small steps. These
explosions could be termed tertiary with the loss of few hundred atoms. It
might be interesting to see if such secondary and tertiary explosion(s)
could be observed experimentally.

\begin{figure}
 
\begin{center}
\includegraphics[width=\linewidth]{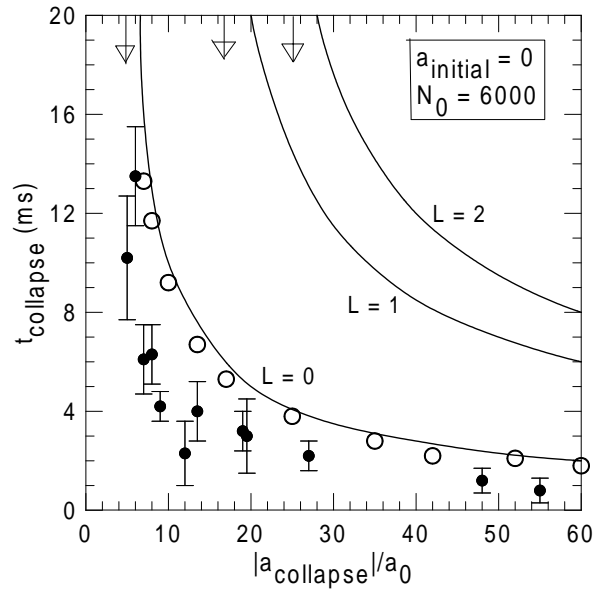}
\end{center}

\caption{The collapse time $t_{\mbox{collapse}}$
vs. $|a_{\mbox{collapse}}|/a_0$
for $a_{\mbox{initial}}=0$ and  $N_0=6000$ for different $L$. Solid circle
with
error bar:
experiment \cite{ex4} for $L=0$;  open circle: axially symmetric
mean-field model of
Ref. \cite{th3} for $L=0$; arrows are the $a_{\mbox{cr}}/a_0$ values,
full line: present theory for different $L$.}

\end{figure}

Donley et al. \cite{ex4} provided a quantitative measurement  of the
variation of
collapse time $t_{\mbox{collapse}}$ with the final scattering length
$a_{\mbox{collapse}}$ for a given 
$a_{\mbox{initial}} =0$, $N_0 =6000$, and   $L=0$.
We calculated this variation using our model   for $L=0, 1$ and 2.
The  $t_{\mbox{collapse}}$ vs. $|a_{\mbox{collapse}}|/a_0$  plots for
$L=0,1,2$ are exhibited  in Fig. 4 and
compared with experimental
data for $L=0$ \cite{ex4} as well as with another calculation using the
mean-field
GP equation in an axially symmetric trap  for $L=0$ \cite{th3}. 
We see $t_{\mbox{collapse}}$ decreases with
$|a_{\mbox{collapse}}|/a_0$ starting from an infinite value at
$|a_{\mbox{collapse}}|=a_{\mbox{cr}}$,   where
$a_{\mbox{cr}}$ is the
minimum value of $|a_{\mbox{collapse}}|$ that leads to collapse and
explosion. The
critical
value $a_{\mbox{cr}}$ increases with $L$ and so does 
$t_{\mbox{collapse}}$ for a fixed  $|a_{\mbox{collapse}}|$.  
For a given $N_0$, a critical value of $n \equiv
n_{\mbox{cr}} $ can be defined
 via $n_{\mbox{cr}} \equiv  N_0
a_{\mbox{cr}}/l$. A necessary condition for collapse is
$N_0|a_{\mbox{collapse}}|/l> n_{\mbox{cr}}$ \cite{sk1}. 
 The value of $n_{\mbox{cr}} $ for a specific case  is
calculated as in Ref. \cite{sk1}. Consequently, $a_{\mbox{cr}}/a_0$ can be
obtained. The value of  $a_{\mbox{cr}}/a_0$ so evaluated for a specific
$L$ is shown by an arrow near the curve for that particular $L$ in
Fig. 4. The $t_{\mbox{collapse}}$ vs. $|a_{\mbox{collapse}}|/a_0$ 
 curves should tend to infinity at the respective arrows and they do so in
Fig. 4. There should not be any collapse for $|a_{\mbox{collapse}}| <
a_{\mbox{cr}}$.

\begin{figure}[!ht]
 
\begin{center}
\includegraphics[width=\linewidth]{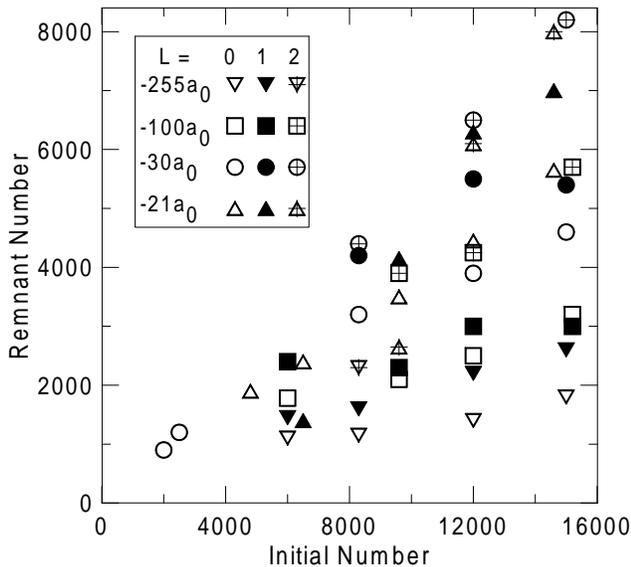}
\end{center}

\caption{Remnant number vs. initial number for  $a_{\mbox{initial}}=7a_0$,
$L=0$, 
$1$ and 2, 
and
different  $a_{\mbox{collapse}}$. 
The
results are represented by different types of  triangle,
circle,  square, and inverted triangle for $L=0, 1,$ and 2 and 
$a_{\mbox{collapse}}=-21a_0, -30a_0, -100a_0 $ and $-255a_0$ as indicated 
in the figure.} 

\end{figure}

Donley et al. \cite{ex4} measured the number of remnant atoms for $L=0$, 
$a_{\mbox{initial}}=7a_0$ and different $N_0$ and $a_{\mbox{collapse}}$
\cite{th4}. We plot the same in Fig. 5 for $L=0, 1$ and 2.  The remnant
number plotted in this figure is the number after the primary
explosion(s)   and not during or after possible secondary
and tertiary
explosions at larger times. In Figs. 1 and 2 the remnant number is
obtained around $t\sim 
20 - 30$ ms and not at few hundred milliseconds. Our results in Fig. 5 
for $L=0$ agree well \cite{th4} with the measurements of Donley et
al. \cite{ex4}. In general the remnant number decreases with
increasing $L$. However, there are some cases where the opposite trend has
been observed in Figs. 1 and 2 and as well as in Fig. 5. For the smallest 
values of $N_0$ in Fig. 5, the condensate remains stable for $L>0$, and
there is no collapse and explosion and hence no 
remnant numbers for $L=1$ and 2.  For certain $N_0$ the results for all
three $L$'s are not plotted as they coincide with other remnant numbers.    
The remnant number in  some cases 
could be much larger than $N_{\mbox{cr}} $ for times on the order of tens
of milliseconds.

\begin{figure}
 
\begin{center}
\includegraphics[width=\linewidth]{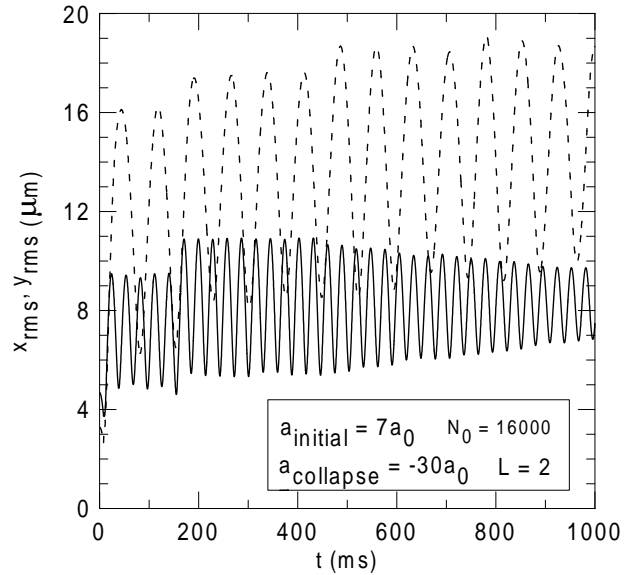}
\end{center}

\caption{The rms sizes $x_{\mbox{rms}}$ (full line) and
$y_{\mbox{rms}}$
(dashed line) 
after the jump in the scattering
length of a BEC of  16000 $^{85}$Rb atoms for $L=2$ from
$a_{\mbox{initial}}=7a_0$
to
$a_{\mbox{collapse}}=-30a_0$
as functions of
time  $t$.}
 
\end{figure}

Donley et al.  \cite{ex4} observed that for $L=0$ the remnant condensate
always 
oscillated in a highly excited collective state with approximate
frequencies $2\nu _{\mbox{axial}}$ and $2\nu _{\mbox{radial}}$ being
predominantly excited. 
 This behavior emerges from the present simulation for all 
values of $L$. To illustrate this  
we plot in Fig. 6 sizes $x_{\mbox{rms}}$ and $y_{\mbox{rms}}$ vs. time for
the condensate after the jump in the scattering length to $-30a_0$ from
$7a_0$ for $N_0=16000 $ and $L=2$. We find a periodic oscillation in
$x_{\mbox{rms}}$ and $y_{\mbox{rms}}$ with frequencies 13.6 Hz
($\simeq 2\nu
_{\mbox{axial}}$)
and 35 Hz ($\simeq 2\nu _{\mbox{radial}}$),
respectively, as observed in experiment.

\section{Conclusion}

In conclusion, we have employed a numerical simulation based on the
solution \cite{sk1}
of the mean-field Gross-Pitaevskii equation \cite{8} with  cylindrical
symmetry  to study the dynamics of collapse and explosion \cite{ex4,th4}
of small
attractive vortex states with $L>0$. The explosion is initiated by a
sudden jump in
the scattering length from a positive to negative value exploiting a
Feshbach resonance \cite{fbna,ex3,fbcs,fbth}.  
 In the GP equation we include a
quintic
three-body nonlinear recombination loss term \cite{th1,th1a,th2,th3} 
that
accounts for the decay
of the strongly attractive condensate.  
The results of the present
simulation are to be considered as an extension of the experiment at JILA
for $L=0$ \cite{ex4}
to small vortex states.

We find  the following features of this dynamics from the present
numerical simulation: (1) The condensate undergoes  collapse
and explosion during a small interval of time of few milliseconds 
and finally stabilizes to a remnant condensate containing a fraction of initial
number of atoms.
The number in the remnant condensate  for
times on the order of tens of milliseconds
can be much larger than the critical
number for
collapse $N_{\mbox{cr}}$ for the same atomic interaction. (2) In some
cases after the primary explosion small secondary and tertiary 
explosions
are observed.  
(3) The explosion takes place during  a decay time of few milliseconds. 
This decay time for rotating Bosenova with vortex ($L>0$) is smaller than 
the same for nonrotating condensate with $L=0$.
(4) The remnant condensate executes radial and
axial oscillations in a highly excited collective state for a long time
with frequencies $2\nu_{\mbox{radial}}$ and $2\nu_{\mbox{axial}}.$ (5)
After the sudden change in the scattering length to a large negative
value, the condensate needs an interval of time $t_ {\mbox{collapse}}$
before it experiences loss via explosion.  The interval  $t_
{\mbox{collapse}}$   increases with $L$ and $a_ {\mbox{collapse}}$.

The simulation of the particle loss in strongly attractive rotating Bosenova,  
with a single axial vortex of small angular momentum
per particle, may stimulate further theoretical
and experimental studies. We have considered small vortex states as they
can be well described by the mean-field GP equation. This will provide a
test for the usefulness of this equation in handling particle loss. 
Otherwise, a similar study with a large  Bose-Einstein
condensed vortex lattice \cite{exp1,expm} is more challenging from both 
experimental and theoretical points.

\acknowledgments

The work is supported in part by the CNPq and FAPESP
of Brazil.

\end{document}